\documentclass[twoside]{article}

\usepackage{lipsum} 

\usepackage[sc]{mathpazo} 
\usepackage[T1]{fontenc} 
\usepackage{microtype} 

\usepackage[hmarginratio=1:1,top=32mm,columnsep=20pt]{geometry} 
\usepackage{multicol} 
\usepackage{hyperref} 

\usepackage[hang, small,labelfont=bf,up,textfont=it,up]{caption} 
\usepackage{booktabs} 
\usepackage{float} 

\usepackage{lettrine} 
\usepackage{paralist} 

\usepackage{abstract} 

\usepackage{titlesec} 
\renewcommand\thesection{\Roman{section}}
\titleformat{\section}[block]{\large\scshape\centering}{\thesection.}{1em}{} 

\usepackage{fancyhdr} 
\title{\vspace{-15mm}\fontsize{24pt}{10pt}\selectfont\textbf{Quantum Gravity on a Quantum Computer?}} 

\author{
\large
\textsc{Achim Kempf\thanks{Talk presented at the International Workshop \it Horizons of Quantum Physics, \rm Taipei, Taiwan, Oct. 14-18, 2012}}\\[2mm] 
\normalsize 
              Department of Applied Mathematics, University of Waterloo, \\
\normalsize 200 University Ave. W., Waterloo N2L 3G1, Ontario, Canada and \\
          \normalsize    Centre for Quantum Computing Technology, Department of Physics\\
\normalsize University of Queensland, St. Lucia, QLD 4072, Australia.\\
 \\ 
\normalsize \href{}{akempf@math.uwaterloo.ca} 
\vspace{-5mm}
}
\date{}

\begin{document}

\maketitle 

\thispagestyle{fancy} 


\begin{abstract}

\noindent 

It is known that EPR-type measurements on spatially separated entangled spin qubits allow one, in principle, to detect curvature. In fact, the spins are not strictly necessary for this purpose since also the vacuum state's spatial entanglement is affected by curvature. Here, we ask if the curvature of spacetime can be expressed entirely in terms of the structure of the spatial entanglement of the vacuum. This opens up the prospect that quantum information techniques could be fully employed in the study of quantum gravity, and that quantum gravity could be simulated on a quantum computer.

\end{abstract}


\section{Introduction}
\label{intro}

The quantization of gravity has proven to be exceptionally difficult \cite{QGtexts}.
Mirroring these difficulties is the fact that quantum computers \cite{nielsen} do not
naturally understand or work with spacetime curvature. If we are to find a way to encode and simulate  spacetime curvature in quantum computers, an interesting lead is the fact that measurements of quantum entanglement, e.g.,
in EPR type experiments, could allow one, in principle, to detect aspects of
spacetime curvature \cite{satellite}. This is because the relative orientation of entangled spins depends on the curvature along their path of travel from creation to detection. 

Since entanglement is an entity that quantum computers do naturally work with, let us here pursue the idea \cite{ak1,ak2} that spacetime curvature might be expressible \it entirely \rm in terms of
entanglement, namely in terms of the entanglement structure of the quantum vacuum. If so, instead of describing spacetime as a manifold with nontrivial curvature, we could equivalently describe spacetime as a quantum field theory with a nontrivial entanglement structure of its vacuum, a description which could help with the quantization of gravity.  Also, using a quantum computer, one could then implement a symbolic representation of a curved manifold by, for example, giving a large number, $N$, of its subsystems (qudits) the same mutual entanglement as the vacuum entanglement between $N$ points of the manifold. The representation is symbolic rather than analogue because the actual spatial arrangement of these qudits within the quantum computer is arbitrary, as only their entanglement matters. In this context, see also \cite{analogues,Weinfurtner}. Ultimately, putting quantum gravity theories on a quantum computer could be useful because quantum gravity theories tend to be mathematically at least as hard as, say, QCD. A quantum computer could therefore help calculating the predictions of a quantum gravity theory efficiently. 

The first task at hand here is no less than to build a suitable bridge between the mathematical frameworks of general relativity and quantum theory. The two frameworks are quite different, with one, namely differential geometry, involving such concepts as curvature, metrics and parallel transports, while the other, functional analysis, involves the rather different concepts of entanglement, Hilbert spaces and spectra.   

Fortunately, there exists a mathematical discipline, spectral geometry, which could be useful here. The field of spectral geometry \cite{haz}, whose origins go back at least to Weyl \cite{Weyl} in the 1910s, addresses a number of inter-related questions. The spectral geometric question of most interest to us here will be `To what extent does the spectrum of the Laplacian on a compact Riemannian manifold determine the metric of the manifold?' For example, if vibrations of a vase are excited, say by touching it with a spoon, how much will the acoustic spectrum reveal about the shape of the vase? 

For completeness only, let us recall other questions addressed in the field of spectral geometry, such as: `To what extent does the spectrum of the Laplacian on a flat domain determine the shape of its boundary?' This includes Kac's iconic question \cite{Kac} `Can one hear the shape of a drum?' Spectral geometry also asks, for example, to what extent the spectrum of a Hamiltonian determines its potential. 

Here, however, our aim is to express spacetime curvature in terms of quantities, such as entanglement and spectra, that are amenable to a quantum theoretic description. To this end, we will pursue the spectral geometry of the type that tries `to hear curvature'. The literature is mostly concerned with highly symmetric manifolds \cite{haz} since spectral geometry is hard due to its nonlinearity. Crucially, among these, also examples of non-isometric manifolds were found that are isospectral, i.e., that possess the same spectra. We will push ahead, nevertheless, to see why in these cases shape cannot be heard and what may be required to make spectral geometry work. 

To this end, we will apply an ultraviolet cutoff and then cut the spectral geometric problem into an iteration of infinitesimal linearized steps that can be treated with powerful methods of perturbation theory. We will find that small curvature changes can indeed be detected from spectral changes (and this process can be iterated) but that for this it is necessary that one `listens' to \it all \rm vibrational degrees of freedom that the manifold possesses, and, crucially, these generally include, beyond scalars, also vectors and tensors. 

We will also find that in order to determine a spacetime's vibrational spectrum, no analog of a `spoon' is needed, since spacetime perpetually rings on its own, due to quantum noise. We will show that the spectrum of the quantum noise can be determined from the correlations of the quantum noise at different points in space. These spatial correlations of the vacuum fluctuations express the fact that the vacuum is a spatially entangled state. What we will find here is that the vacuum state is not only entangled. Indeed, its spatial entanglement structure could encode all of the curvature of the spacetime.   

This new perspective then leads us to questions such as: What is the Einstein action in terms of the structure of the entanglement of the vacuum state? Assuming the existence of a natural Planck scale cutoff, as we here are, what is the baseline density of degrees of freedom in spacetime? Can curvature be viewed as a modulation of this density of degrees of freedom? 

\section{Can one `hear' a manifold's curvature in its quantum noise?}

Let us consider a 2-dimensional compact Riemannian manifold without boundary, or simply a `manifold' for short, embedded in $I\!\!R^3$, perhaps in the shape of the skin of a museum dinosaur. Let us try to capture its shape without the usual tools, such as parallel transport and affine connection. To this end, let us sprinkle the manifold with $N$ uniformly randomly chosen points and let us record the $N\times N$ matrix, $G$, of their pairwise distances $G_{nm}$. Assume that after a museum fire one is left with only this matrix $G$. Can one reconstruct the shape of the dinosaur by gluing together rods of lengths $G_{nm}$ to form a type of scaffolding? Generically, this should work, as the rods will generally not fit together for any other shape but the dinosaur's. Of course, the finiteness of $N$ implies that there is an ultraviolet cutoff on the short distance structures that are captured by $G$. But $N$ could be chosen arbitrarily large. The fact that pairwise distances encode curvature was remarked upon by Einstein, who credits Helmholtz, but the general idea goes back to Gau\ss.  

In order to obtain a quantum version, let us recall that the vacuum state of any quantum field is a spatially entangled state and that fluctuations of the field at two positions, say $x$ and $x'$, are correspondingly correlated. For example, in the simple case of a free, neutral and massless scalar field, $\hat{\phi}$, this is measured by the equal-time 2-point correlation function:
\begin{equation}
 G(x,x') = \langle 0\vert \hat{\phi}(x)\hat{\phi}(x')\vert 0 \rangle 
\end{equation}
Since $G(x,x')$ is a function that decays when the distance between the points $x$ and $x'$ is increased, it 
can serve as a substitute for length measurements. Let us therefore now consider a compact Riemannian manifold without boundary. The compactness is an infrared cutoff that ensures that the spectra are discrete. This could be a Lorentzian signature spacetime at a fixed time or it could be a euclidean signature (i.e., Wick rotated) spacetime.  We sprinkle the spacetime with $N$  uniformly randomly chosen points $\{x_n\}_{n=1}^N$ and we record the matrix, $G$, whose elements are the mutual 2-point correlators, $G_{nm}=G(x_n,x_m)$. Since these matrix elements encode distances, we expect that they encode the shape of the spacetime up to the cutoff scale $V/N$.  

This would mean (we will address caveats below) that one can determine the shape of a spacetime up to a cutoff scale by recording (in an ensemble measurement) its quantum vacuum noise at $N$ points $\{x_n\}_{n=1}^N$ and by then calculating their two-point correlators $G_{nm}$. To see intuitively why this could be possible, recall that the vacuum is spatially entangled because the local field oscillators are coupled through covariant derivatives, while the covariant derivatives depend on the curvature. Thus, the vacuum state is not the product state of the ground state of the local oscillators but is a curvature-dependent entangled state. We recall also that the associated vacuum entanglement entropy is implicated in holography and the entropy of black holes \cite{sred}. 

So far, our entanglement-based method for capturing the curvature of the spacetime suffers, however, from the arbitrariness of the sprinkling of the $N$ points. If we choose another sprinkling of $N$ points, say $\{\tilde{x}_n\}_{n=1}^N$, we obtain new matrix elements $G(\tilde{x}_n,\tilde{x}_m) = \langle 0\vert \hat{\phi}(\tilde{x}_n)\hat{\phi}(\tilde{x}_m)\vert 0 \rangle$. They describe the same curved manifold down to the cutoff scale. This leads us to ask: What in $G$ is encoding the shape of the manifold and what in $G$ is merely a function of the sprinkling? This question is analogous to the question of what in the metric $g_{\mu\nu}$ is describing the curvature of a manifold and what in the metric is merely a gauge degree of freedom that reflects a choice of coordinates. In the case of the metric, the two kinds of degrees of freedom are notoriously hard to separate. Can we do better in the case of the matrix $G$?   

To see that this is the case, let us recall that the propagator, or 2-point correlator, $G$, is the inverse of the kinetic term in the action. In the case of a massless field on a euclidean signature spacetime,
$$
\Delta G(x,x') = \delta(x-x')
$$ 
where $\Delta$ is the Laplace operator on square integrable functions on the curved manifold. Therefore, $G$ is the inverse operator to the Laplacian and, using functional analytic notation, we can express $G(x,x')$ as a matrix element:
\begin{equation}
 G(x,x') = (x,\Delta^{-1}, x') 
\end{equation}
Here, $\vert x)$ is a position eigenvector on the manifold. Let us now assume a spatially covariant ultraviolet cutoff by restricting the space of fields to the space spanned by only the first $N$ eigenvectors $\{\vert b_n)\}_{n=1}^N$ of $\Delta$, with eigenfunctions $b_n(x)=(b_n,x)$.  Using the projector $P=\sum_{n=1}^N \vert b_n)( b_n\vert$:
\begin{equation}
 G(x,x') = (x, P\Delta^{-1}P,  x')  
\end{equation}
Therefore, the matrix elements $G(x_n,x_m)=(x_n, P\Delta^{-1}P, x_m)$ simply represent the UV cut off operator $G= P\Delta^{-1}P$ in the generally non-orthonormal basis $\{P\vert x_n)\}_{n=1}^N$. To change to a new sprinkling of points, $\{\tilde{x}_n\}$, merely amounts to representing $G$ in a new basis: $G(\tilde{x}_n,\tilde{x}_m)=(\tilde{x}_n, P\Delta^{-1}P, \tilde{x}_m)$.

We conclude that the sprinkling-independent information in $G$ is basis independent information in $G$, and that is the spectrum of $G$. As a consequence, given a sprinkling of points and the corresponding matrix $G_{nm}$, we can now distil sprinkling-independent curvature information out of the matrix, namely by calculating its eigenvalues. 

But the eigenvalues of $G$ are the inverses of the eigenvalues of $\Delta$. We therefore arrive at the conjecture that the spectrum of the Laplacian of a manifold, and therefore the vibrational spectrum of the manifold, determines its shape. This happens to be exactly one of the questions that has been pursued in the field of spectral geometry for a long time, although we arrived at this conjecture here from an entirely new perspective. 

But does the conjecture hold true? The literature on spectral geometry has some tantalizing positive indications in this regard, but it does also provide counter examples \cite{haz}, such as certain high-dimensional tori that possess the same spectrum but are non-isometric. Rather than giving up at this point, let us now revisit spectral geometry to find out exactly why the conjecture as stated cannot hold. The answer will be related to the fact that curvature is tensorial rather than scalar in dimensions larger than two. Therefore, for the spectrum to determine the shape it will be necessary to consider the spectra of the Laplacians arising from 2-point correlators of quantum fluctuations not only of scalar fields but also of vector and suitable tensor fields. 

\section{Revisiting spectral geometry}

The `shape' or metric of a manifold fully determines the spectrum of its Laplace operator but it does so in a highly nonlinear way. This makes it hard to determine if or where the map from metric to spectrum may be invertible.  In the mathematical literature, the problem has usually been simplified by considering highly symmetric situations. Spacetime, however, has no exact symmetry. In order to be able to obtain results for the spectral geometry of realistic spacetimes, i.e., of generic manifolds without symmetries, let us adopt a new strategy, which is to iterate linearized spectral geometry \cite{ak2}.  

To this end, let us assume that both, a generically-shaped manifold and the spectrum of its Laplacian are known. We then consider a slight perturbation of the shape of the manifold. It will map into a proportionally small change of the spectrum. Is this linear map invertible? If it is invertible, we can then iterate such small reconstructions of shape to reconstruct even large shape changes. When given only the spectrum of a manifold, we can start with an arbitrary manifold of the same global topology and then change its shape in small steps such that its spectrum incrementally approaches that of the target manifold's spectrum. Varying the starting manifold then allows one to probe the uniqueness of the shape for a given target spectrum.  

Concretely, let us describe a small change in the shape of a manifold through a small-amplitude scalar function, $f$, say $g_{\mu\nu}\rightarrow (1+f)g_{\mu\nu}$. We assume finite vision and finite hearing, i.e., we assume again that the spectrum of $\Delta$ is ultraviolet cut off at the $N$'th eigenvalue. Expanding $f$ in the Laplacian's eigenbasis, $f(x)=\sum_{n=1}^Nf_n b_n(x)$, we can express the small shape changes through the $N$ coefficients $f_n$. Small changes of shape lead to proportional changes in the Laplacian: $\Delta \rightarrow \Delta + \sum_{n=1}^N \tilde{\Delta}_n f_n$. The change of the eigenvalues $\lambda_n$ of the Laplacian, $\lambda_n\rightarrow \lambda_n+\mu_n$, is then given by $N$ coefficients $\mu_n$, which we can calculate using conventional first order perturbation theory
$$
\mu_n = \sum_{n=1}^N S_{nm} f_m
$$
with $S_{nm} = (b_n, \tilde{\Delta}_m b_n)$. 
Let us define $r= \mbox{rank}(S)$. By observing how the $N$ eigenvalues change, we can reconstruct changes of shape from within an $r$-dimensional space of shape changes. In fact, the rank of $S$ is at most $N-1$. This is because for the compact Riemannian manifolds without boundary that we  consider here, zero is always an eigenvalue \cite{Rosenberg}, i.e., $\lambda_1\equiv 0$ and therefore $\mu_1=0$ and $\det(S)=0$. This costs one degree of freedom in the spectrum. On the other hand, we may assume one shape degree of freedom as given, say the overall size. The so-reduced $S$ is a $(N-1)\times(N-1)$ matrix with generic entries and we may expect its determinant to be nonzero, so that it is of full rank, $r=N-1$. 

If so, does this mean that we can calculate from any small change of sound the small change of shape that caused it?
In \cite{ABK}, it was found that the answer is indeed yes, in the case where the manifolds are 2-dimensional, star-shaped and triangulated. In fact, it was found, using explicit numerical algorithms, that not only small changes of shape but also large changes of shape could be iteratively reconstructed from spectral information alone. 

For manifolds of dimensions larger than two, the situation is more subtle, however. This is because in the above discussion we neglected an important aspect of manifolds of dimension larger than two, namely that their curvature degrees of freedom are not scalar. This means that small changes of shape  cannot be expressed simply through a scalar function $f$. As a consequence, there are many more degrees of freedom of shape than just the $\{f_n\}$. 

Namely, any small change to the metric is generally given by a symmetric covariant 2-tensor $h_{\mu\nu}$. Any such a tensor can be expanded, as in studies of cosmological structure formation \cite{Liddle}, into effectively scalar, vector and tensor contributions, analogously to how vector fields are decomposed into a gradient part, which is effectively scalar, and a curl part. These contributions can be expanded in the eigenbases $\{b^{(s)}_n\},\{b^{(v)}_n\}$ and $\{b^{(t)}_n\}$ of Laplace operators on scalars, vectors and symmetric covariant 2-tensors, for eigenvalues up to an ultraviolet cutoff, which we may call $\Lambda$. Now the total number of eigenfunctions again matches the total number of eigenvalues of these Laplace operators, $N=N^{(s)}+N^{(v)}+N^{(t)}$, counting degeneracies, so that the total $S$ is quadratic. It should be very interesting to study the rank of $S$, which we conjecture to be close to $N$ (as above, zero is always an eigenvalue).    

It is intuitive that in order to hear all possible small shape changes of a manifold we need to listen to all types of shape perturbations that the manifold supports, which are in general scalar, vectorial and tensorial.  The fact that the spectra of these types of waves carry independent shape information is analogous to the situation in seismology. There, independent geological information can be obtained from analyzing both scalar pressure and tensorial shear waves, since they possess different propagation properties (e.g., shear waves cannot penetrate the liquid interior). In the case of spacetime, one would record the spatially entangled fluctuations of scalar, vector and tensor metric perturbations, or equivalently, the spatially entangled fluctuations of scalar, vector and tensor matter fields. These could be fields of fundamental particles such as Higgs or photons for scalars and vectors, or suitable composite particles, which could also provide the spin-2 case (down to the scale of the size of the composite particles).   

\section{Conclusions}

We investigated whether spacetime curvature might be expressible in terms of its `vibrational' spectrum. We found that this could work, but only if we record the spectra of spacetime with respect to all the types of waves that it supports. Also, we should caution that our methods only cover the reconstruction of small shape changes from the corresponding small changes of these spectra. This may be sufficient for studies of perturbative quantum gravity. But it is nontrivial to iterate such small perturbative steps  to arrive at the full reconstruction of a curved manifold from its spectrum, even assuming no changes in the global topology. Namely, there is no guarantee of uniqueness. For example, a nonperturbative phenomenon is that the spectra of Laplacians are invariant under a mirroring of a manifold. This indicates that the spectrum of waves corresponding to handed particles should be needed to hear a manifold's handedness.  

Let us now return to the original question of whether spacetime curvature could be expressed entirely in terms of entanglement, so that, for example, it could be naturally encoded in a quantum computer. To this end, we went beyond merely expressing spacetime curvature in terms of the manifolds' spectra. We also used the fact that these spectra are present in the perpetual ringing of spacetime due to quantum fluctuations. Namely, we found that  the quantum noise spectra can be distilled from samples of the spatial correlations in the quantum vacuum noise. We used the 2-point correlators of quantum fluctuations at $N$ randomly sprinkled points to reconstruct the spacetime similarly to how we used the distances of points sprinkled on a museum dinosaur to reconstruct its skin. 

We found, however, that to this end the spectra of the Laplacians on scalars, vectors and symmetric covariant two-tensors are needed, which means that we should record the 2-point functions of quantum fluctuations of scalar, vector and tensor fields. An interesting message here is that not only do the correlations of the vacuum quantum fluctuations at two points depend on the curvature of spacetime. Also, we now know that the quantum fluctuations of the different types of fields carry different information about the spacetime curvature. Not only is the spatial structure of the entanglement of the scalar, vector and tensor vacua dependent on the spacetime curvature, but also, vice versa, the spatial entanglement structure of the full vacuum may well encode the spacetime curvature completely (up to discrete information such as parity). Concretely, it may be possible to model a spacetime with curvature in a quantum computer for example by encoding the points that we sprinkle on the manifold as q-dits whose entanglement structure matches that of the spacetime.  

\section{Outlook}
There are many open questions on the horizon, of course. Let us consider a few and discuss speculative ideas to address them. For example, recall that we are here describing Riemannian manifolds, not Lorentzian manifolds. This means that we can only treat either spacelike hypersurfaces of a full Lorentzian-signature spacetime, or the full four-dimensional spacetime after Wick rotation. In the former case, we will need to develop the presently only spatially covariant approach into a fully spacetime covariant approach, e.g., by the cutting of the spectrum of the d'Alembertian rather than that of the Laplacian. For work in this direction, see \cite{aidanrob}. In this context, see also \cite{Landi,Weinfurtner}. 

In the latter case, i.e., when we apply our methods to euclidean signature spacetimes, an immediate question is: what is the expression of the Einstein action in terms of spectra? 
One might expect the answer to be complicated. Namely, one might expect that one first needs to use the spectrum to calculate the metric, then calculate the Ricci scalar which is finally integrated over the spacetime to obtain the Einstein action.  In fact, the Einstein action, up to small covariant corrections, can be expressed very easily in spectral terms. To this end, we can use a result of spectral geometry \cite{Hawking}:
\begin{equation}
\label{hawking}
N=\frac{1}{16\pi^2}\int d^4x~\sqrt{\vert g\vert}\left\{\frac{\Lambda^2}{2}+\frac{\Lambda}{6}~R + \Lambda^0 O(R^2) +...\right\}
\end{equation}
Here, $O(R^2)$ denotes terms that involve two or more factors of the Riemann curvature tensor or its covariant derivatives. $N$ is the number of eigenvalues of the scalar Laplacian which are below the cutoff $\Lambda$. Let us choose for $\Lambda$ the Planck value $\Lambda=L_{Pl}^{-2}$. We can then write the gravity action as:
\begin{eqnarray*}
S_{grav} & = &\frac{6\pi}{16\pi^2}\int d^4x~\sqrt{\vert g\vert}\left\{\frac{1}{2 L_{Pl}^4}+\frac{1}{6L_{Pl}^2}~R + L_{Pl}^0 O(R^2) +...\right\}\\
  & = & 6\pi~N\\
  & = & 6\pi~\mbox{Tr} (1)
  \end{eqnarray*}
This action, $S_{grav}$, contains the Einstein term $S_E=(16\pi L_{Pl}^2)^{-1}\int d^4x~\sqrt{\vert g\vert}R$. The higher order correction terms are beyond observability. This type of terms arises from quantum fluctuations in any effective gravity action since they are not forbidden by any known symmetries.  The fact that the gravity action $S_{grav}$ can be written as the trace of the identity is interesting because every field's action can be written as a trace. For example, the action of a free Klein Gordon field is $S_{KG}=\mbox{Tr}\left((\Delta -m^2)\vert \phi)(\phi\vert\right)$. Gravity's contribution to the action can, therefore, be seen as the leading constant in an expansion. 

We notice also that we obtain a contribution to the cosmological constant that is, as usual, at least 120 orders of magnitude too large. Our new perspective does not offer an explanation for how its gravitational effect is offset. But we do obtain a new interpretation of the role of this term. To see this, consider first the case without curvature. In this case, from the expression for the manifold's volume, $V= \int d^4x~\sqrt{\vert g \vert}$, we obtain from Eq.\ref{hawking} that the density $\rho_s(x)$ of points sprinkled on the manifold, $N/V$, is given by:
$$
\rho_s = N/V = \frac{\Lambda^2}{32\pi^2}
$$
Therefore, $\Lambda$ determines the baseline density of the scalar degrees of freedom of the manifold. We also see that curvature locally modulates that density: 
$$\rho_s(x) = \frac{\Lambda^2}{32\pi^2} + \frac{\Lambda}{96\pi^2}R + ...$$
This leads us to ask whether, vice versa, the local curvature tensor could be calculated from the local density of degrees of freedom. It is clear that the knowledge of the manifold's spatially varying density of scalar degrees of freedom, $\rho_s(x)$, cannot suffice, since curvature is tensorial. But it may be sufficient to use the generalizations of Eq.\ref{hawking} for the Laplacians on vectors and tensors. These may also allow one to linearly combine a gravity action in which the contributions to the cosmological constant suitably cancel. 

We have here outlined a path to expressing curvature in terms of spectra and entanglement, in view of possibly working with gravity on a quantum computer. To this end our strategy is to first express the spacetime curvature in terms of the spectra of the quantum noise on the spacetime. Then, these spectra of the quantum noise can be expressed in terms of the correlations of field fluctuations at a discrete set of points. The finite density of these sample points implies that we obtain the spectra only up to an ultraviolet cutoff. These spectra can therefore determine the shape of the manifold only down to wrinkles at the scale set by the cutoff. Given the ultraviolet cutoff, manifolds are indistinguishable that differ only in eigenvalues that are higher than the cutoff. 

This situation, an ultraviolet cutoff and reconstruction of a shape from discrete samples taken at the density set by the ultraviolet cutoff, is reminiscent of Shannon sampling theory.  Recall that Shannon sampling theory \cite{shannon} is in ubiquitous use in all of signal processing and communication engineering. This is because it  is at the heart of information theory where it establishes the equivalence of continuous and discrete representations of information. Concretely, Shannon sampling theory holds that any function $f$ that does not contain frequencies larger than some `bandlimit' $\Omega$, i.e., for which there exists an $\tilde{f}$ so that
$$f(x) = \int_{-\Omega}^{\Omega}d\omega \tilde{f}(\omega)~e^{ i\omega x}$$  
can be \it perfectly \rm reconstructed from any set of its amplitude samples $f(x_n)$,  if their average spacing is at most $\pi/\Omega$, see \cite{beurling}. For example, if the samples are taken equidistantly, $x_n=n \pi/\Omega$, the reconstruction formula reads \cite{shannon}:
$$
f(x) = \sum_n f(x_n)~\frac{\sin( (x-x_n)\Omega)}{ (x-x_n)\Omega}
$$
Sampling theory can be generalized from signals to wave functions and fields and it is in fact deeply related \cite{ak-firstshannoninqft} to minimum length uncertainty principles that have been suggested to arise from quantum gravity and string theory \cite{Garay,Hossenfelder}. Hilbert space representations of such uncertainty principles were first developed in \cite{ak-ucr}. Using these, it was shown \cite{ak-firstshannoninqft} that a minimum length uncertainty principle always implies and is implied by bandlimitation of the wave functions. The minimum position uncertainty is in one to one correspondence to the bandlimit. 

The bandlimitation of fields can be generalized to quantum fields over curved spacetime, by applying an ultraviolet cutoff to its Laplacian's spectrum \cite{ak-firstcurved}. Notice that this cutoff is spatially covariant because the Laplacian's spectrum is scalar. Covariant sampling theory offers the opportunity to reconcile two competing views of spacetime at the Planck scale, namely that spacetime is either fundamentally continuous or discrete. Using sampling theory, spacetime with this ultraviolet cutoff can be viewed as being simultaneously continuous and discrete, in the same mathematical way that information is both discrete and continuous simultaneously \cite{ak1,ak2}. Fields, equations of motion and actions can all be written, equivalently, as living on a continuous manifold or on any one of the sufficiently dense discretizations. (The fact that none of the discretizations is singled out means also that, unlike with a fixed lattice, no spacetime symmetries need to be broken.) Also, sampling theory would replace Wheeler's hard-to-describe concept of spacetime foam with straightforward quantum uncertainty. 

Our study in this paper can be viewed as an approach to developing a Shannon sampling theory for curvature. Here, the aim is not merely to reconstruct a function on a known curved background spacetime from samples of its amplitudes. This would be essentially ordinary Shannon sampling for fields. Instead, we here take discrete samples of quantum noise to then reconstruct the metric of the spacetime. The central idea is that the entanglement structure of the scalar, vector and tensor vacuum could fully encode the curvature of spacetime. The hope on the quantum horizon is that if this program can be carried to full fruition, more quantum information theoretical tools, including quantum computers, will become available to theoretical studies of quantum gravity itself. 
$$$$
\bf Acknowledgements: \rm AK acknowledges support from the Discovery and Canada Research Chairs programmes of NSERC, as well as the kind hospitality during his sabbatical stay at the University of Queensland where this work was completed.



\end{document}